\documentclass[nofootinbib,showpacs,preprintnumbers,amsmath,amssymb,floatfix]
{revtex4-2}
\usepackage{graphicx}
\begin{document}


\title{Rise of the DIS structure function $F_L$ at small $x$ caused by double-logarithmic contributions}

\vspace*{0.3 cm}

\author{B.I.~Ermolaev}
\affiliation{Ioffe Physico-Technical Institute, 194021
 St.Petersburg, Russia}
\author{S.I.~Troyan}
\affiliation{St.Petersburg Institute of Nuclear Physics, 188300
Gatchina, Russia}

\begin{abstract}
We present calculation of $F_L$ in the double-logarithmic  approximation and demonstrate that the synergic effect of the factor $1/x$
from the $\alpha_s^2$-order and the steep $x$-dependence of the totally resummed double logarithmic contributions of higher orders ensures
the power-like rise of $F_L$ at small $x$ and arbitrary $Q^2$.
\end{abstract}


\maketitle

\section{Introduction}

Theoretical investigation of the DIS structure function $F_L (x, Q^2)$ (and other DIS structure functions) in the context of
perturbative QCD began with calculations in the fixed orders in $\alpha_s$. First, there were calculations in the Born approximation,
then more involved first-loop and second-loop  calculations (see Refs.~\cite{flor}-\cite{kotik})
followed by the third-order results\cite{moch3}. All fixed-order calculations showed that $F_L$ decreases at small $x$.
Alternative approach to study $F_L$ was applying all-order resummations. In the first place, $F_L$  was studied with DGLAP\cite{dglap}
and its NLO modifications.  In addition, there are
approaches where DGLAP is combined with
BFKL\cite{bfkl}, see e.g. Ref.~\cite{kwiech, ball}.
Besides, there are calculations in the literature based on the dipole model, see Refs.~\cite{kovch, lusz}. This
approach was used in the global analysis of experimental data in Ref.~\cite{arm}.
Let us notice that Ref.~\cite{zfitter}
contains detailed bibliography on this issue.

Applying DGLAP to studying $F_L$ is model-independent. However according to Ref.~\cite{zfitter}, neither LO DGLAP
nor the NLO DGLAP modifications ensure the needed rise of $F_L$ at small $x$ and disagree with experimental data
at small $Q^2$, which sounds quite natural because DGLAP by definition is not supposed to be used in
the region of small $Q^2$. The modifications of DGLAP in Refs.~\cite{kwiech, ball} are
based on treating BFKL as a small-$x$ input for the DGLAP equations.
The approach of Ref.~\cite{abf} treats the Pomeron
intercept as a parameter fixed from experiment.

In this paper we present an alternative approach to calculate $F_L$:
total resummation of double-logarithmic (DL) contributions to $F_L$, accounting for both logarithms of $x$
and $Q^2$.
The method we use is self-consistent and does not involve any models.
We modify the approach which we used in Ref.~\cite{etf1} to calculate $F_1$ in the
Double-Logarithmic Approximation (DLA). This approach has nothing in common with the BFKL equation and its
ensuing modifications. Indeed, instead of summing leading logarithms, i.e. the contributions
$\sim (1/x) \left(\alpha_s \ln (1/x)\right)$, which is the BFKL domain,
we sum the DL contributions $\sim \alpha_s \ln^2 (1/x)$ as well as the DL of $Q^2$. Because of the absence of the factor $1/x$ such
contributions were  commonly neglected by the HEP community for a long time. However, it has recently
 been proved in Ref.~\cite{etf1} that the DL contribution to Pomeron is not
less important than the BFKL contribution.

We calculate $F_L$ in DLA with constructing
and solving Infra-Red Evolution Equations (IREEs). As is well-known, the IREE approach was suggested by L.N.~Lipatov\cite{kl}. It proved to be a simple and
efficient instrument (see e.g. the overviews in Ref.~\cite{egtg1sum}) for calculating many objects in QCD and Standard Model.
Constructing and solving IREEs, we obtain general solutions. In order to specify them
  one has to define the starting point (input) for IREEs.
Conventionally in the IREE technology the Born contributions have been chosen as the inputs .
However, $F_L = 0$ in the Born approximation, so
the input has to be chosen anew. We suggest that the the second-loop expression for $F_L$ can play the role of the input
and arrive thereby to explicit expressions for perturbative components of $F_L$.
We demonstrate that the total resummation
of DL contributions together with the factor $1/x$ appearing in the $\alpha^2_s$-order provide $F_L$ with the rise
at small $x$.

We start with considering $F_L$ in the large-$Q^2$ kinematic region

\begin{equation}\label{kin}
Q^2 > \mu^2,
\end{equation}
with $\mu$ being a mass scale. Then we present a generalization of our results to small $Q^2$. The scale $\mu$ is often associated with the factorization scale.
The value of $\mu$ is arbitrary\footnote{for specifying $\mu$ on basis of Principle of Minimal Sensitivity\cite{pms} see Ref.~\cite{egtg1sum}} except the requirement $\mu > \Lambda_{QCD}$
to guarantee applicability of perturbative QCD.

Our paper is organized as follows:
In Sect.~II we introduce definitions and notations, then remind how to calculate $F_L$
through auxiliary invariant functions.
Calculations of  $F_L$ in the $\alpha_s^2$-order are considered in Sect.~III.
We represent them in the way convenient for analysis of contributions from higher loops.
  Then we explain how to realize our strategy:
combining the non-logarithmic
results from the $\alpha_s^2$-order with double-logarithmic (DL) contributions from higher-order graphs. Total resummation of DL contributions to
$F_L$ is done in Sect.~IV through constructing and solving IREEs. IREEs control both $x$ and $Q^2$ -evolutions of $F_L$ from the starting point.
 Specifying the input is done in Sect.~V. In Sect.~VI we present explicit expressions for
leading small-$x$ contributions to perturbative components of $F_L$. To make clearly seen the rise of $F_L$ at small $x$ we
present
the small-$x$ asymptotics of $F_L$.
After that we compare our results for  $F_L$ at small $x$ with the ones predicted by approaches involving BFKL.
Then we consider the generalization of our results on $F_L$ in region (\ref{kin}) to the small-$Q^2$ region.
Finally, Sect.~VII is for concluding remarks.

\section{Calculating $F_L$ through auxiliary amplitudes}

The most convenient way   to calculate $F_{1,2}$ and $F_L$ in Perturbative QCD is
the use of auxiliary invariant amplitudes. Below we remind how this approach works.
The unpolarized part of the hadronic tensor describing the lepton-hadron DIS is

\begin{eqnarray}\label{wgen}
W_{\mu\nu} (p,q) &=& \left(-g_{\mu\nu} + \frac{q_{\mu}q_{\nu}}{q^2}\right) F_1 + \frac{1}{pq}
\left(p_{\mu} - q_{\mu}\frac{pq}{q^2}\right)\left(p_{\nu} - q_{\nu}\frac{pq}{q^2}\right) F_2
\end{eqnarray}
and each of $F_1,F_2$ depends on $Q^2$ and $x = Q^2/w$, with $Q^2 = - q^2$ and $w = 2pq$.
It is convenient to represent $F^{(q,g)}_{1,2}$ through
auxiliary amplitudes $A$ and $B$ which are the convolutions of the tensor
$W^{(q,g)}_{\mu\nu}$ with $g_{\mu \nu}$ and $p_{\mu} p_{\nu}$:

\begin{eqnarray}\label{defa}
-A &\equiv& g_{\mu\nu} W_{\mu\nu}
= 3 F_1 + \frac{F_2}{2x}  + O(p^2),
\end{eqnarray}

\begin{eqnarray}\label{defb}
B \equiv \frac{p_{\mu}p_{\nu}}{pq} W_{\mu\nu}
&=& -\frac{1}{2x} F_1 + \frac{1}{4x^2} F_2 + O(p^2),
\end{eqnarray}
where we use the standard notatons $x = - q^2/w = Q^2/w$, $w = 2pq$.
Neglecting terms $\sim p^2$,
we express $F_{1,2}$ through $A$ and $B$:

\begin{eqnarray}\label{fab}
F_1 &=& \frac{A}{2} + x B,
\\ \nonumber
F_2
&=&
2xF_1 + 4x^2 B,
\end{eqnarray}
so that

\begin{equation}\label{fldef}
F_L = F_2 - 2x F_1 = 4 x^2 B.
\end{equation}

Each of $F_1, F_2$
includes both perturbative and non-perturbative contributions. According to the QCD factorization concept, these contributions
can be separated.
\begin{figure}
\includegraphics[width=.6\textwidth]{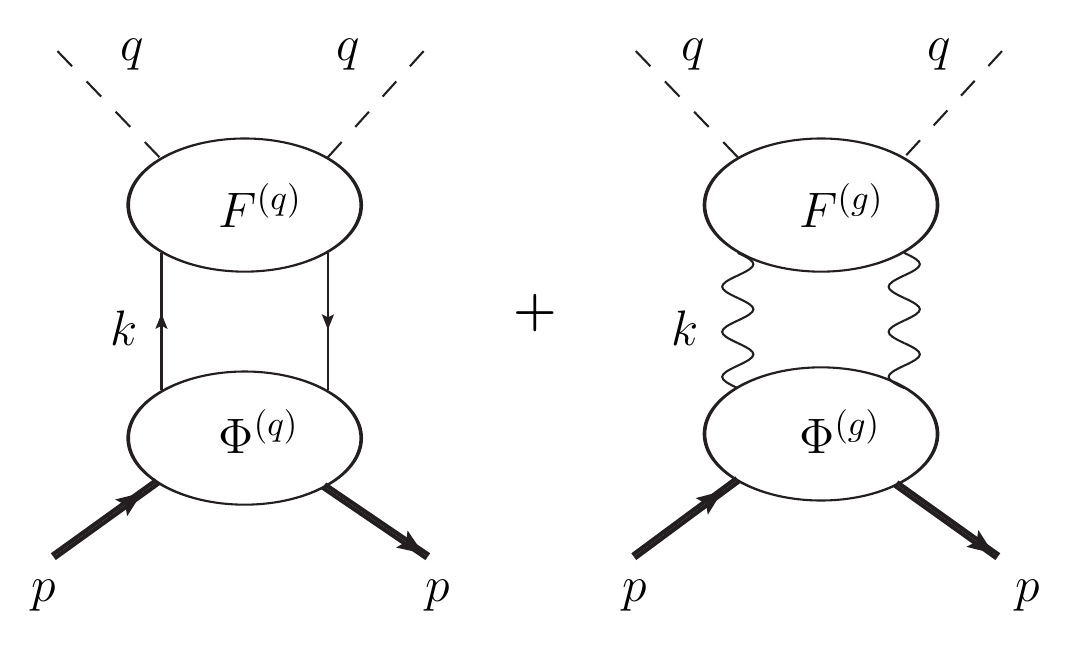}
\caption{\label{flfig1} QCD factorization for DIS structure functions. Dashed lines denote
virtual photons. The upper blobs describe DIS off partons.
The straight (waved) vertical lines denote virtual quarks (gluons) . The lowest
blobs correspond to initial parton distributions in the hadrons.
$F^{(q,g)}$ is a generic notation for perturbative components of $F_1^{(q,g)}$,
$F_2^{(q,g)}$ and $F_L^{(q,g)}$.}
\end{figure}
In scenario of the single-parton scattering,  $F_1, F_2$
can be represented in any available form of QCD factorization through the following
convolutions (see Fig.~\ref{flfig1}):

\begin{equation}\label{fact}
F_1 = F_1^{(q)} \otimes \Phi_{(q)} + F_1^{(g)} \otimes \Phi_{(g)}, ~~~~F_2 = F_2^{(q)} \otimes \Phi_{(q)} + F_2^{(g)} \otimes \Phi_{(g)},
\end{equation}
where $\Phi_{1,2}^{(q, g)}$ stand for initial parton distributions whereas $F_1^{(q,g)}, F_2^{(q,g)}$ are perturbative components of
the structure functions $F$ and $F_2$ respectively.
 The superscripts $q (g)$ in Eq,~(\ref{fact}) mean that the initial partons in the perturbative Compton scattering are quarks (gluons).
 The DIS off the partons is parameterized by the same way as Eq.~(\ref{wgen}):

\begin{eqnarray}\label{wpert}
W^{(q,g)}_{\mu\nu} (p,q) &=& \left(-g_{\mu\nu} + \frac{q_{\mu}q_{\nu}}{q^2}\right) F^{(q,g)}_1 + \frac{1}{pq}
\left(p_{\mu} - q_{\mu}\frac{pq}{q^2}\right)\left(p_{\nu} - q_{\nu}\frac{pq}{q^2}\right) F^{(q,g)}_2,
\end{eqnarray}
with $p$ denoting the initial parton momentum.  Throughout the paper we will neglect virtualities $p^2$, presuming
the initial partons to be nearly on-shell. Introducing the auxiliary amplitudes $A^{(q,g)}$ and $B^{(q,g)}$ similarly
to Eqs.~(\ref{defa},\ref{defb}), one can express $F_1^{(q,g)}$ and $F_2^{(q,g)}$ in terms of $A^{(q,g)}$ and $B^{(q,g)}$
so that

\begin{equation}\label{flqgdef}
F^{(q,g)}_L = F^{(q,g)}_2 - 2x F^{(q,g)}_1 = 4 x^2 B^{(q,g)},
\end{equation}
with

\begin{equation}\label{bqgdef}
B^{(q,g)} = \frac{p_{\mu} p_{ \nu}}{pq} W^{(q,g)}_{\mu\nu}.
\end{equation}

Applying (\ref{flqgdef},\ref{bqgdef}) to $W^{(q,g)}_{\mu\nu}$ in the Born and  first-loop approximation
yields (see Refs.~\cite{flor}-\cite{kotik}) that $F^{(q)}_L = F^{(g)}_L = 0$ in the Born approximation whereas the first-loop results are:

\begin{equation}\label{fl1}
\left( F_L^{(q)}\right)_{(1)} = \frac{2\alpha_s}{\pi} C_F  x^2 ,~~ \left( F_L^{(g)}\right)_{(1)} =
\frac{4\alpha_s}{\pi}  n_f x^2 (1-x).
\end{equation}

Eq.~(\ref{fl1}) suggests that $F_L$ should decrease $\sim x^2$ at $x \to 0$. However, the second-loop
results exhibit a slower decrease.

\section{Leading contributions to $B$ in the second-loop approximation}

The second loop brings a radical change to the small-$x$
behaviour of $B$ compared to the first-loop result. Namely, there appear contributions $\sim 1/x$ in
contrast to logarithmic dependence of $B$ in the first loop.
Such contributions were calculated in Ref.~\cite{neerv}. Nevertheless, we prefer to repeat these
calculations in order to represent the results
in the way convenient for applying to
the total resummation of higher loops in DLA. Doing so, we account for the leading contributions only. Throughout the paper we use the Feynman gauge for virtual gluons.

\begin{figure}
\includegraphics[width=.8\textwidth]{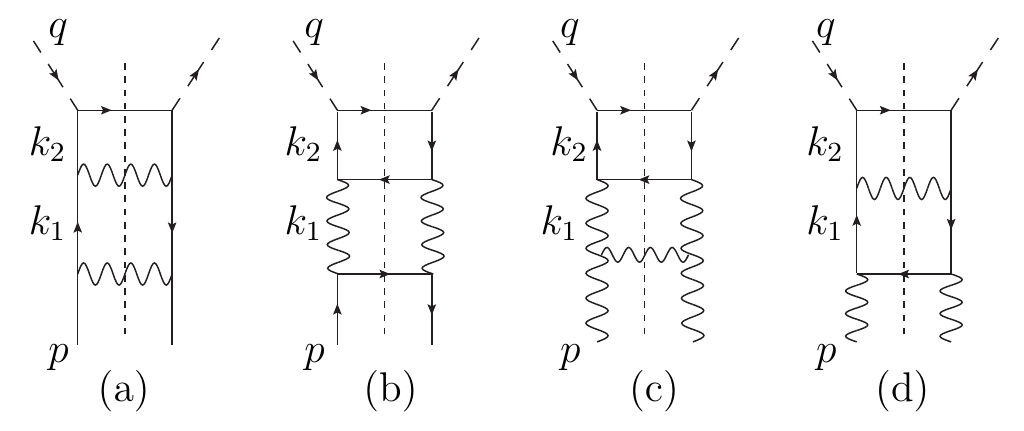}
\caption{\label{flfig2} Ladder graphs for $F_1^{(q,g)}$, $F_2^{(q,g)}$ and $F_L^{(q,g)}$ in the
second-loop approximation.
Graphs (a) and (b) correspond to DIS off quarks and graphs (c) and (d)
are for DIS off gluons.}
\end{figure}

In the first place we consider ladder graphs contributing to $B$,
The ladder graphs contributing to $W_{\mu \nu}$ in the $\alpha_s^2$-order are depicted in Fig.~\ref{flfig2}.
Graphs (a) and (b) correspond to  DIS
off quarks whereas graphs (c) and (d) are for DIS off gluons.
Calculations in the small-$x$ kinematics are simpler when the Sudakov variables\cite{sud} are used. In terms of them, momenta $k_i$
of virtual partons
are parameterized as follows:

\begin{equation}\label{sud}
k_i = \alpha_i q' + \beta_i p' + k_{i\perp},
\end{equation}
where $q'$ and $p'$ are the massless (light-cone) momenta made of momenta $p$ and $q$:

\begin{equation}\label{pqprime}
p' = p - q (p^2/w) \approx p, ~~~ q^{\prime} = q - p (q^2/w) = q + x p.
\end{equation}

In Eq.~(\ref{pqprime}) $q$ denotes the virtual photon momentum while $p$ is momentum of the initial parton.
We remind that we presume that $p^2$ is small, so we will neglect it throughout the paper.
Invariants involving $k_i$ look as follows in terms of the Sudakov invariants:

\begin{eqnarray}\label{sudinv}
k^2_i &=& w \alpha_i \beta_i - k^2_{i \perp} = w (\alpha_i \beta_i - z_i),~~2pk_i = w \alpha_i,~~ 2qk_i = w (\beta_i - x \alpha_i),
\\ \nonumber
2k_i k_j &=& w (\alpha_i - \alpha_j)(\beta_i - \beta_j) -  k^2_{i \perp}- k^2_{j \perp}
= w \left((\alpha_i - \alpha_j)(\beta_i - \beta_j) - z_i - z_j \right).
\end{eqnarray}
We have introduced in Eq.~(\ref{sudinv}) dimensionless variables $z_{i,j}$ defined as follows:

\begin{equation}\label{zi}
z_i = k^2_{i \perp}/w.
\end{equation}

\subsection{Contributions to $B$ for DIS off quarks}

We start with calculating the second-loop
contribution $B^{(a)}_q$  of the two-loop ladder graph (a) in Fig.~\ref{flfig2} to $B$ for DIS off quarks. It is given by the following expression:

\begin{eqnarray}\label{ba}
B^{(2a)} = C^{(2a)}_q \chi_2 w \int d \alpha_{1,2} d \beta_{1,2} d k^2_{1,2 \perp} \frac{N^{(2a)}}{k^2_1 k^2_1 k^2_2 k^2_2}
\delta \left((q + k_2)^2\right) \delta \left((k_1 - k_2)^2\right) \delta \left((p-k_1)^2\right),
\end{eqnarray}
where $C^{(2a)}_q = C^2_F$,

\begin{equation}\label{chi2}
\chi_2 =  \frac{\alpha^2_s}{8 \pi}
\end{equation}
and

\begin{eqnarray}\label{na}
N^{(2a)}  &=& \frac{1}{2} Tr \left[\hat{p}\gamma_{\lambda_1}\hat{k}_1 \gamma_{\lambda_2}\hat{k}_2 \hat{p}
(\hat{q}+ \hat{k}_2) \hat{p}\hat{k}_2 \gamma_{\lambda_2}\hat{k}_1 \gamma_{\lambda_1}\right]
= 2 k^2_1 Tr \left[\hat{k}_2 (\hat{k}_1 -\hat{p}) \hat{k}_2 \hat{p}
(\hat{q}+ \hat{k}_2) \hat{p}  \right]
\\ \nonumber
&=& 2 k^2_1 (w + 2pk_2) Tr \left[\hat{k}_2 (\hat{k}_1 -\hat{p}) \hat{k}_2 \hat{p} \right].
\end{eqnarray}

We represent it as the sum of $N^{(2a)}_1$ and $N^{(2a)}_2$:

\begin{equation}\label{nal2}
N^{(2a)} = N^{(2a)}_1 + N^{(2a)}_2
\end{equation}

with
\begin{eqnarray}\label{na1}
 N^{(2a)}_1
=  - 4 k^2_1  \left((2pk_2)^3 + w (2pk_2)^2\right)
\end{eqnarray}
and
\begin{eqnarray}\label{na2}
N^{(2a)}_2 =  4k^2_1\left[ \left(k^2_1 + k^2_2\right) \left((2pk_2)^2 + w (2pk_2)\right) - k^2_1 k^2_2 (2pk_2) - w k^2_1 k^2_2  \right],
\end{eqnarray}

In Eqs.~(\ref{na1},\ref{na2}) we have used  the quark density matrix

\begin{equation}\label{rhoq}
\hat{\rho}(p) = \frac{1}{2} \hat{p}
\end{equation}

and made use of the $\delta$-functions of Eq.~(\ref{ba}).
They yield that  $2k_1k_2 = k^2_1 + k^2_2$ and $2pk_1 = k^2_1$. It turns out that the
leading contributions comes from $N^{{2a}}_1$, so first of all we consider it.
Throughout the paper we will use dimensionless variables $z_{1,2}$ instead of $k^2_{1,2 \perp}$:

\begin{equation}\label{z12}
z_1 = k^2_{1 \perp}/w,~~z_2 = k^2_{2 \perp}/w,~~z = z_1 + z_2.
\end{equation}

It is also convenient to use the variable $l$ defined as follows:

\begin{equation}\label{l}
l = \beta_1 - \beta_2.
\end{equation}

Using the $\delta$-functions to integrate (\ref{ba}) over $\alpha_{1,2}$ and $\beta_2$ and replacing
$ N^{(2a)}$ by $ N^{(2a)}_1$ we are left with three more integrations:

 \begin{eqnarray}\label{b2qsud}
B^{(2a)}
&\approx&   C^{(2a)}_q \chi_2 \int^1_{\lambda}  \frac{dz_1}{z_1} \int^1_{\lambda} \frac{dz_2}{z^2_2} \int^1_z dl
\left[- \frac{z^3}{ l^2 (l + \eta)^2} + \frac{z^2}{l (l + \eta)^2} \right],
\end{eqnarray}
with $\eta$ defined as follows:

\begin{equation}\label{eta}
\eta = \frac{z (x + z_2)}{z_2}.
\end{equation}

Details of calculation in Eq.~(\ref{b2qsud}) can be found in Appendix A. Let us remind that
throughout this paper we focus on the small-$x$ region.  The most important contributions in Eq.~(\ref{b2qsud}) at small $x$ are $\sim 1/x$.
Retaining them only and integrating (\ref{b2qsud}) with logarithmic accuracy, we arrive at

\begin{equation} \label{b4alead}
B^{(2a)}  \approx
 C^{(2a)}_q \gamma^{(2)} ~x^{-1},
\end{equation}
with
\begin{equation}\label{lambda}
\gamma^{(2)} = 4 \chi_2  \rho \ln2,
\end{equation}
where $\chi_2$ is defined in (\ref{chi2}) and

\begin{equation}\label{rho}
\rho = \ln (w/\mu^2),
\end{equation}

with $\mu$ being an infrared cut-off.
Contribution to $B$ of graph (b) in Fig.~\ref{flfig2} is given by the following expression:

\begin{eqnarray}\label{bb}
B^{(2b)} = C^{(2b)}_q
\chi_2 w \int d \alpha_{1,2} d \beta_{1,2} d k^2_{1,2 \perp} \frac{N^{(2b)}}{k^2_1 k^2_1 k^2_2 k^2_2}
\delta \left((q + k_2)^2\right) \delta \left((k_1 - k_2)^2\right) \delta \left((p-k_1)^2\right),
\end{eqnarray}
where $C^{(2b)}_q = n_f C_F $ and

\begin{equation}\label{nb}
N^{(2b)} = p_{\mu} p_{\nu} Tr \left[\gamma_{\nu} \left(\hat{q} + \hat{k}_2\right)\gamma_{\mu}\hat{k}_2 \gamma_{\lambda^{\prime}}
\left(\hat{k}_1 - \hat{k}_2\right) \gamma_{\sigma^{\prime}} \hat{k}_2 \right] \left(p_{\lambda^{\prime}} k_{1 \sigma^{\prime}}
+ k_{1 \lambda^{\prime}} p_{1 \sigma^{\prime}}\right).
\end{equation}

Apart from the factor $C^{(2b)}_q$, the integrand in Eq.~(\ref{bb}) coincides with the integrand of Eq.~(\ref{ba}), so we
obtain the same leading contribution:

\begin{equation}\label{b4blead}
B^{(2b)} \approx C^{(2b)}_q ~x^{-1} \gamma^{(2)},
\end{equation}
where $\gamma^{(2)}$ is given by Eq.~(\ref{lambda}).
Our analysis of non-ladder graphs shows that they do not bring the factor $1/x$ because they do not
contain $(k^2_2)^2$ in denominators. Therefore, the total leading contribution $B^{(2)}_q$ to $B_q$ in
the second loop is

\begin{equation}\label{b2qlead}
B^{(2)}_q = \left(C^{(2a)}_q + C^{(2b)}_q\right)  \gamma^{(2)}~x^{-1} \equiv C^{(2)}_q  \gamma^{(2)}~x^{-1}.
\end{equation}

Now let us consider some important technical details concerning Eqs.~(\ref{b4alead}) (the same reasoning holds for Eq.~(\ref{b4blead})).
This result stems from the terms in Eq.~(\ref{na}) where momenta $k_2$  are coupled with the external momenta $p$ and $q$.
The other terms in Eq.~(\ref{na}) (i.e. the ones $\sim k^2_2, k_1 k_2$)  either cancel $k^2_2$ in the denominator of Eq.~(\ref{ba}), preventing
appearance of the factor $1/x$, or cancel $1/k^2_1$, killing $\ln w$. Hence, the first step to  calculate the trace in Eq.~(\ref{na}) can be reducing
the trace
down to $Tr [\hat{p}\hat{k}_2\hat{p}\hat{k}_2]$.
Obviously, it corresponds to neglecting the factor $2pk_1$ in $\hat{k}_1 \hat{p} \hat{k}_1$:

\begin{equation}\label{kpk1}
\hat{k}_1 \hat{p} \hat{k}_1 = 2pk_1 \hat{k}_1 - k^2_1 \hat{p} \approx - k^2_1 \hat{p}.
\end{equation}
This observation allows us to develop a strategy to select most important contributions to $B$ in arbitrary orders
in $\alpha_s$. In other words, the non-singlet component of $F_L$ can be calculated in DLA in the straightforward way,
without evolution equations.

\subsection{Contributions to $B$ for DIS off gluons}

 The second-loop contributions to the DIS off the initial gluon correspond to
 the ladder
 graphs (c,d) in Fig.~\ref{flfig2}. We calculate their joint contribution $B_g$
 to $F_L$. Obviously, the contribution of graph (c) is

\begin{equation}\label{bc}
B^{(2c)} = C^{(2c)}_g \chi^{(2)} \int dz_{1,2} d \beta_{1,2} d \alpha_{1,2}
\frac{N^{(2c)}}{k^2_1 k^2_1 k^2_2 k^2_2} \delta \left((p - k_1)^2\right) \delta \left((k_1 - k_2)^2\right),
\delta \left((q + k_2)^2\right)
\end{equation}
where $\chi^{(2)}$ is defined in Eq.~(\ref{chi2}) and $C^{(2c)}_g = n_f N$.
%
%
The numerator $N^{(2c)}$ is defined as follows:

\begin{equation}\label{nc}
N^{(2c)} = p_{\mu} p_{\nu} Tr \left[\gamma_{\nu} \left(\hat{q} + \hat{k}_2\right)\gamma_{\mu}\hat{k}_2 \gamma_{\lambda^{\prime}}
\left(\hat{k}_1 - \hat{k}_2\right) \gamma_{\sigma^{\prime}} \hat{k}_2 \right] H_{\lambda^{\prime} \sigma^{\prime}},
\end{equation}
with

\begin{equation}\label{habgen}
H_{\lambda^{\prime} \sigma^{\prime}} = H_{\lambda^{\prime} \sigma^{\prime} \lambda \sigma} \rho_{\lambda \sigma}.
\end{equation}

In Eq.~(\ref{habgen}) the notation $H_{\lambda^{\prime} \sigma^{\prime} \lambda \sigma}$ stands for the ladder gluon rung while
$\rho_{\lambda \sigma}$ denotes the gluon density matrix for the initial gluons which we treat as slightly virtual:

\begin{eqnarray}\label{htot}
H_{\lambda^{\prime} \sigma^{\prime} \lambda \sigma} = -
\left[(2k_1 - p)_{\lambda}g_{\lambda^{\prime} \tau} + (2p - k_1)_{\lambda^{\prime}} g_{\lambda \tau} + (- k_1 - p)_{\tau}g_{\lambda^{\prime} \lambda} \right]
\\ \nonumber
\left[(2k_1 - p)_{\sigma}g_{\sigma^{\prime} \tau} + (2p - k_1)_{\sigma^{\prime}} g_{\sigma \tau}
+ (- k_1 - p)_{\tau}g_{\beta \sigma} \right].
\end{eqnarray}

The terms $\sim p_{\lambda}, p_{\sigma}$ in
(\ref{htot}) can be dropped because of the
gauge invariance.
We use the Feynman  gauge for the initial gluons:
\begin{equation}\label{rhog}
\rho_{\lambda \sigma} = - \frac{1}{2} g_{\lambda \sigma}.
\end{equation}

As a result we obtain

\begin{equation}\label{hab}
H_{\lambda^{\prime} \sigma^{\prime}} = 8 p_{\lambda^{\prime}} p_{\sigma^{\prime}} -
4 (p_{\lambda^{\prime}} k_{1 \sigma^{\prime}} + k_{1 \lambda^{\prime}} p_{\sigma^{\prime}}) +
2 k_{1 \lambda^{\prime}} k_{1 \sigma^{\prime}}
+3 g_{\lambda^{\prime} \sigma^{\prime}} k^2_1.
\end{equation}
We have used in the last term of (\ref{hab}) that $2pk_1 \approx k^2_1$. DL contributions to the gluon ladder come
from the kinematics where $\lambda^{\prime} \in R_L, \sigma^{\prime} \in R_T$ or vice versa (the symbols $R_L$ and $R_T$
denote the longitudinal and transverse momentum spaces respectively). Therefore, the leading term
in (\ref{htot}) in DLA is

\begin{equation}\label{hdl}
H^{DL}_{\lambda^{\prime} \sigma^{\prime}} = - 4 \left(p_{\lambda^{\prime}} k_{1 \sigma^{\prime}} + k_{1 \lambda^{\prime}} p_{\sigma^{\prime}}\right)
\end{equation}
while $2 k_{1 \lambda^{\prime}} k_{1 \sigma^{\prime}}$ brings corrections to it. The first term in (\ref{hab})
contain the longitudinal momenta only and the last term vanishes at $\lambda^{\prime} \neq \sigma^{\prime}$.
Substituting (\ref{hdl}) in (\ref{nc}) we obtain

\begin{eqnarray}\label{ncdl}
N^{(2c)}_g &=& Tr \left[\hat{p} \left(\hat{q} + \hat{k}_2\right)\hat{p}\hat{k}_2 \hat{p}
\left(\hat{k}_1 - \hat{k}_2\right) \hat{k}_{1 \perp} \hat{k}_2 \right] +
Tr \left[\hat{p} \left(\hat{q} + \hat{k}_2\right)\hat{p}\hat{k}_2 \hat{k}_{1 \perp}
\left(\hat{k}_1 - \hat{k}_2\right) \hat{p} \hat{k}_2 \right]
\\ \nonumber
&\approx& Tr \left[\hat{p}\hat{q}\hat{p}\hat{k}_2 \hat{p}
\left(\hat{k}_1 - \hat{k}_2\right) \hat{k}_{1 \perp} \hat{k}_2 \right] +
Tr \left[\hat{p}\hat{q}\hat{p}\hat{k}_2 \hat{k}_{1 \perp}
\left(\hat{k}_1 - \hat{k}_2\right) \hat{p} \hat{k}_2 \right]
\\ \nonumber
&=& w  Tr \left[\hat{p}\hat{k}_2 \hat{p}
\left(\hat{k}_1 - \hat{k}_2\right) \hat{k}_{1 \perp} \hat{k}_2 \right] +
w Tr \left[\hat{p}\hat{k}_2 \hat{k}_{1 \perp}
\left(\hat{k}_1 - \hat{k}_2\right) \hat{p} \hat{k}_2 \right]
\\ \nonumber
&=& w 2pk_2 Tr \left[\hat{p}\left(\hat{k}_1 - \hat{k}_2\right)\hat{k}_{1 \perp} \hat{k}_2\right]
+  w 2pk_2 Tr \left[\hat{p} \hat{k}_2 \hat{k}_{1 \perp}\left(\hat{k}_1 - \hat{k}_2\right) \right]
\end{eqnarray}

Retaining in (\ref{ncdl}) the terms $\sim (pk_2)^2$ and $\sim (pk_2)^3$, we obtain the leading contribution to $N^{DL}_g$:

\begin{equation}\label{nglead}
N^{(2c)}_g \approx 4 (w + 2pk_2) (2pk_2)^2 k^2_{1 \perp}
\end{equation}
which coincides with $N^{(2a)}_1$.
Substituting $N^{(2c)}_g$ in (\ref{bc}), representing $B_g$  as

\begin{equation}\label{bgi}
B^{(2c)}_g  = C^{(2)}_g \chi^{(2)} I_g
\end{equation}
and then integrating over $\alpha_2$,  we arrive at

\begin{eqnarray}\label{icdef}
I^{(c)}_g = \int^1_{\lambda} \frac{dz_1}{z_1} \int^1_{\lambda} \frac{dz_2}{z^2_2} \int^1_z d l
\left[- \frac{z^3}{l^2(l + \eta)^2} + \frac{z^2}{l(l + \eta)^2}\right]
\end{eqnarray}
with $z,z_{1,2}, l$ and $\eta$ defined in Eqs.~(\ref{z}) and (\ref{eta}) respectively. The integral in
Eq.~(\ref{icdef}) coincides with the integral bringing  the leading contribution to $B^{(2a)}_q$ in (\ref{b2qsud}).
obtained for the quark ladder graph and calculated in Appendix A. So, we arrive at the leading
contribution to $B$:

\begin{equation}\label{b4clead}
B^{(2c)}_g  \approx
  C^{(2c)}_g x^{-1} \gamma^{(2)},
\end{equation}
with $\gamma^{(2)}$ defined in Eq.~(\ref{lambda}).
Now calculate contribution  $B^{(2d)}$   to $B^{(2)}_g$ of graph (d) in Fig.~\ref{flfig2}. It is given by the following expression:

\begin{eqnarray}\label{bdgen}
B^{(2d)}_g = - C^{(2d)}_g  \chi_2 \int dz_{1,2} d \beta_{1,2} d \alpha_{1,2} \frac{N^{(d)}_g}{k^2_1 k^2_1 k^2_2 k^2_2}
\delta \left((p - k_1)^2\right) \delta \left((k_1 - k_2)^2\right),
\delta \left((q + k_2)^2\right)
\end{eqnarray}
where $\chi_2$ is defined in Eq.~(\ref{chi2}) and $C^{(2d)}_g = n_f C_F$.

\begin{eqnarray}\label{ndgen}
N^{(2d)}_g &=& \frac{1}{2} Tr \left[\hat{p}\left(\hat{q} + \hat{k}_2 \right) \gamma_{\rho}\hat{k}_1
\gamma_{\lambda}\left(\hat{k}_1  -  \hat{p} \right)\gamma_{\lambda} \hat{k}_1 \gamma_{\rho}\hat{k}_2\right]
\\ \nonumber
&=& 2 \left(w + 2pk_2 \right) Tr \left[\hat{p}\hat{k}_2 \hat{k}_{1} \left(\hat{k}_1  -  \hat{p} \right)
\hat{k}_1 \hat{k}_2\right],
\end{eqnarray}
where we have used the gluon density matrix of Eq.~(\ref{rhog}). Retaining the terms with $pk_2$ and neglecting other terms
containing $k_2$, we obtain

\begin{eqnarray}\label{ndgdl}
N^{(2d)}_g &\approx&  2 \left(w + 2pk_2 \right) k^2_1 Tr \left[\hat{p}\hat{k}_2 \hat{p} \hat{k}_2 \right]
= 4 \left(w + 2pk_2 \right) (2pk_2)^2 k^2_1.
\end{eqnarray}

substituting Eq.~(\ref{ndgdl}) in (\ref{bdgen}), introducing variables $l, z_{1,2}$, then accounting for the $\delta$-functions, we arrive at

\begin{eqnarray}\label{bdgdl}
B^{(2d)} \approx C^{(2d)}_g \chi_2 \int_{\lambda}^1  \frac{dz_1}{z_1} \int_{\lambda}^1  \frac{dz_2}{z^2}
\int_z^1 \frac{dl }{ (l + \eta)^2}  \left[- \frac{z^3}{l^2} + \frac{z^2}{l}\right],
\end{eqnarray}
with $\eta$ defined in Eq.~(\ref{eta}). Comparison of (\ref{bdgdl}) with Eq.~(\ref{b2qsud}) shows that
apart of the colour factors the leading contribution,
$B^{(2d)}_L$ to $B$ coincides with $B^{({2a})}_q$:

\begin{equation}\label{b4dlead}
B^{(2d)}_g  = C^{(2d)}_g  \gamma^{(2)}~x^{-1}.
\end{equation}

Therefore, the total leading contribution $B^{(2)}_g$ to $B_g$ in
the second loop is

\begin{equation}\label{b2glead}
B^{(2)}_g = \left(C^{(2c)}_g + C^{(2d)}_g\right)  \gamma^{(2)}~x^{-1} \equiv C^{(2)}_g  \gamma^{(2)}~x^{-1}.
\end{equation}

Eqs.~(\ref{b4alead}, \ref{b4blead}, \ref{b4clead}) and (\ref{b4dlead}) demonstrate explicitly  that
the only difference between leading contributions of all ladder graphs in Fig.~\ref{flfig2} is different color factors.
Combining Eqs.~(\ref{b2qlead},\ref{b2glead}) with (\ref{bqgdef}) demonstrate that $F_L$ in the
$\alpha^2_s$-order decreases
at $x \to 0$ slower than the first-order result (\ref{fl1}). Nevertheless, there are no growth of $F_L$ in the $\alpha^2_s$-order
and in the $\alpha^3_s$-order as shown in Ref.~\cite{moch3}. It suggests that only all-order resummations
can provide $F_L$ with some growth.

\subsection{Remark on the scale of $\alpha_s$}

The factor $\gamma^{(2)}$ defined in Eq.~(\ref{lambda}) involves the QCD coupling $\alpha_s$  treated as a constant
because of complexity of the two-loop calculations. However, one cannot implement
the expressions for $B^{(2)}_{q,g}$ in Eqs.~(\ref{b2qlead}, \ref{b2glead})
until the scale of $\alpha_s$ has been specified.
The adequate parametrization of $\alpha_s$
for processes in the Regge kinematics was obtained in Ref.~\cite{egtalpha} but it cannot be used in $B^{(2)}_{q,g}$
because the leading contributions there come from the kinematics which is rather hard than Regge.
For this reason, we suggest using  in $B^{(2)}_{q,g}$
the standard DGLAP parametrization $\alpha_s = \alpha_s (Q^2)$.

\subsection{Remark on leading contributions of the ladder graphs in higher loops}

Contribution $B_q^{(n)}$ of the quark ladder graph to $B$ in the $n^{th}$ order of the perturbative expansion can be written as
follows:

\begin{eqnarray}\label{bqngen}
B_q^{(n)} &=& \chi_n C^n_F w^{n-1} \int
d k^2_{1 \perp}...d k^2_{n \perp} d \alpha_1... d \alpha_n d \beta_1...d \beta_n  \frac{N_q^{(n)}}{k^2_1 k^2_1 k^2_2... k^2_n}
\\ \nonumber
&~&\delta \left((q + k_n)^2\right) \delta \left((k_n - k_{n-1})^2\right)... \delta \left((p-k_1)^2\right),
\end{eqnarray}
with
\begin{equation}\label{chinsud}
\chi_n =
 2e^2 \left(- \frac{\alpha_s }{2 \pi^2} \frac{\pi}{2}\right)^n = 2e^2 \left(- \frac{\alpha_s }{4 \pi}\right)^n .
\end{equation}

and

\begin{eqnarray}\label{nqngen}
N_q^{(n)} &=& \frac{1}{2} Tr \left[\gamma_{\lambda_1}\hat{k}_1...\gamma_{\lambda_{n-1}}\hat{k}_{n-1}\gamma_{\lambda_{n-1}}\hat{k}_n\gamma_{\lambda_n}\hat{k}_n \hat{p}\left(\hat{q} + \hat{k}_n\right)\hat{p}\hat{k}_n \gamma_{\lambda_n}\hat{k}_{n-1}\gamma_{\lambda_{n-1}}...\hat{k}_1\gamma_{\lambda_1} \hat{p}\right]
\\ \nonumber
&=& -(w + 2pk_n) Tr \left[\hat{k}_1...\gamma_{\lambda_{n-1}}\hat{k}_{n-1}\gamma_{\lambda_{n-1}}\hat{k}_n\gamma_{\lambda_n}\hat{k}_n \hat{p}\hat{k}_n \gamma_{\lambda_n}\hat{k}_{n-1}\gamma_{\lambda_{n-1}}...\hat{k}_1 \hat{p}\right].
\end{eqnarray}

We have used in (\ref{nqngen}) the quark density matrix given by Eq.~(\ref{rhoq}). We are going to calculate $B_q^{(n)}$ in DLA.
In order to select appropriate contributions in the trace in (\ref{nqngen}),
we generalize the approximation of  Eq.~(\ref{kpk1}) to $k_i$, with $i = 1,2,.., n-1$:

\begin{equation}\label{kpki}
\hat{k}_i \hat{p} \hat{k}_i = 2pk_i \hat{k}_i - k^2_i \hat{p} \approx - k^2_i \hat{p}.
\end{equation}

Doing so we arrive at the DL contribution $N_q^{DL}$:

\begin{eqnarray}\label{nqn}
N_q^{DL} &=& (-2)^{n-1} k^2_1...k^2_{n-1} (w + 2pk_n) Tr [\hat{p}\hat{k}_n \hat{p} \hat{k}_n]
\\ \nonumber
&\approx& 2^{n-1} k^2_{1 \perp}...k^2_{n-1 \perp} Tr [\hat{p}\hat{k}_n \hat{p} \hat{k}_n].
\end{eqnarray}

Substituting (\ref{nqngen}), we arrive at $B_q^{(n)}$ in DLA. The integration region in DLA was found in
\cite{ggfl}:

\begin{eqnarray}\label{ggfl}
\beta_1 \gg \beta_2 \gg ... \gg \beta_n,
\\ \nonumber
\frac{k^2_{1 \perp}}{\beta_1} \ll \frac{k^2_{2 \perp}}{\beta_2} \ll... \ll \frac{k^2_{n-2 \perp}}{\beta_{n-2}}.
\end{eqnarray}

Integrations over momenta $k_1,..., k_{n-2}$ in the region (\ref{ggfl}) yield DL contributions
whereas integration over $k_n, k_{n-1}$ yields the factor $1/x$. Integration over $k_n, k_{n-1}$  is not restricted by Eq.~(\ref{ggfl}) but
runs over the whole phase space. As is known\cite{nest}, contributions of non-ladder graphs cancel each other in DLA.
Such a straightforward approach is comparatively simple
for purely quark ladders (e.g., for non-singlet structure functions)  but becomes too complex for calculating singlets where the quark rungs are mixed with gluon ones. It
is more
practical to implement evolution equations in this case.

\subsection{Remark on contributions of non-ladder graphs}

Our analysis of the non-ladder graphs $\sim \alpha^2_s$ shows that they do not yield the factor $1/x$ and because of that they can be
neglected. The technical the reason of their smallness is that they do not yield $(k^2_2)^2$ in denominators.  At the same time,
non-ladder graphs are essential in higher loops ($\sim \alpha_s^n$, with $n > 2$).
They should be accounted for because they bring DL contributions. However, as long as $\alpha_s$ is treated as a constant,
DL contributions of the non-ladder graphs cancel each other\cite{nest} and therefore they are essential at running $\alpha_s$ only.

\section{Calculating $B_q$ and $B_g$ in DLA}

We calculate $B_q$ and $B_g$ with constructing and solving IREEs for it.
 Constructing
 IREEs in the DIS context was explained in many our papers. For instance, IREEs for the DIS structure function $F_1$
 can be found in \cite{etf1}; the overview of the technical details can be found in Ref.~\cite{egtg1sum}.
 The essence of this approach is first to introduce a IR cut-off $\mu$ to regulate IR divergences of the graphs contributing
 to $B_{q,g}$ in higher loops\footnote{ We use the mass scale $\mu$  of Eq.~(\ref{kin})
 as an IR cut-off for simplicity reason, in order to avoid introducing extra
 parameters.}.  Once such cut-off has been introduced, amplitudes $B_{q,g}$ become $\mu$-dependent and
 tracing their evolution with respect to $\mu$ allows one to construct IREEs.
 The IREE technology involves the IR cut-off which restricts from below transverse
 momenta of virtual partons and exploits the fact that that DL contributions ofWe calculate $B_q$ and $B_g$ with constructing and solving IREEs for it.
 Constructing
 IREEs in the DIS context was explained in many our papers. For instance, IREEs for the DIS structure function $F_1$
 can be found in \cite{etf1}; the overview of the technical details can be found in Ref.~\cite{egtg1sum}.
 The essence of this approach is first to introduce a IR cut-off $\mu$ to regulate IR divergences of the graphs contributing
 to $B_{q,g}$ in higher loops.  Once such cut-off has been introduced, amplitudes $B_{q,g}$ become $\mu$-dependent and
 tracing their evolution with respect to $\mu$ allows one to construct IREEs.
 The IREE technology involves the IR cut-off which restricts from below transverse
 momenta of virtual partons and exploits the fact that that DL contributions of
 the partons with minimal   $k_{\perp}$ can be factorized.

 The IREEs for $B_{q,g}$ take a simpler form when the Mellin
 transform has been used. We are going to calculate dependence of $B_{q,g}$ on both $w$ and $Q^2$
 but the standard parametrization $B_{q,g} = B_{q,g} (x, Q^2/\mu^2)$ leaves the $w$-dependence to be
 $\mu$-independent, so as a result we cannot trace it within the IREE technology.
 Because of that we replace $x$ by the $\mu$-dependent argument $w/\mu^2$, arriving at the parametrization
 $B_{q,g} = B_{q,g} \left(w/\mu^2, Q^2/\mu^2\right)$. We stress that this replacement is purely
 technical detail and the standard parametrization will be restored
 automatically in final expressions for $B_{q,g}$.
 For the present, we write the Mellin transform for $B_{q,g}$  as follows:

 \begin{eqnarray}\label{mellin}
 B_{q,g} \left(w/\mu^2, Q^2/\mu^2\right) &=& \int_{- \imath \infty}^{\imath \infty} \frac{d \omega}{2 \pi \imath} \left(w/\mu^2\right)^{\omega} f_{q,g} \left(w/\mu^2, Q^2/\mu^2\right).
 \end{eqnarray}

As usually, the integration line runs to the right of the
 rightmost singularity of $f_{q,g}$. The transform inverse to Eq.~(\ref{mellin}) is

 \begin{equation}\label{invmellin}
 f_{q,g} (\omega, Q^2/\mu^2) =
 \int_{\mu^2}^{\infty} \frac{d w}{w} \left(w/\mu^2\right)^{- \omega}~B_{q,g} (w/\mu^2, Q^2/\mu^2).
 \end{equation}

 Throughout the paper we will address $f_{q,g}$ as Mellin amplitudes.
 The same form for Mellin transforms we used in Ref.~\cite{etf1} for amplitudes $A_{q,g}$.
 It is convenient to use beyond the Born approximation the logarithmic variables $\rho$ defined in Eq.~(\ref{rho})
 and $y$ defined as follows:

 \begin{equation}\label{rhoy}
 y =  \ln \left(Q^2/\mu^2\right).
 \end{equation}

IREEs for amplitudes $A_{q,g}$ were obtained in Ref.~\cite{etf1} and IREEs for amplitudes
 $B_{q,g}$ are absolutely the same, so we do not derive them here and only briefly comment on them.
 IREEs for $B_{q,g}$ in the $\omega$-space look as follows:

\begin{eqnarray}\label{fqgeqs}
 \partial f_q (\omega, y)/\partial y  &=&
 \left[ - \omega + h_{qq}(\omega) \right] f_q (\omega, y)+ f_g (\omega, y) h_{gq}(\omega),
 \\ \nonumber
  \partial f_g (\omega, y)/\partial y &=& f_q (\omega, y) h_{qg}(\omega) +
  \left[ - \omega + h_{gg}(\omega) \right]f_g (\omega, y),
 \end{eqnarray}
with $h_{qq}, h_{gq}, h_{qg}, h_{gg}$ being auxiliary amplitudes describing parton-parton scattering in DLA.
They can be found in \cite{etf1}.  In addition, explicit expressions for $h_{ik}$ (with $i,k = q,g$) can be found in Appendix B.
One can see that Eqs.~(\ref{fqgeqs}) exhibit a certain similarity to the DGLAP equations. Indeed, the l.h.s.
of Eqs.~(\ref{fqgeqs})
are the derivatives with respect to $\ln Q^2$.
Very soon we will demonstrate that the role of the terms $\sim \omega$ in the r.h.s. of (\ref{fqgeqs}) is to convert
the factor $\left(w/\mu^2\right)^{\omega}$ into $x^{- \omega}$.
The remaining difference between
Eqs.~(\ref{fqgeqs}) and DGLAP equations is that all anomalous dimensions $h_{ik}$ in Eqs.~(\ref{fqgeqs}) are calculated in DLA,
i.e. they contain  contributions $\sim \alpha_s^{1 + n}/\omega^{1 + 2n}$ to all orders in $\alpha_s$
whereas the DGLAP equations operate with the anomalous dimensions calculated in several fixed orders in $\alpha_s$.
For instance, the most singular terms in the LO DGLAP they are $\sim \alpha_s/\omega$
while NLO DGLAP involves more singular terms.
General solution to Eq.~(\ref{fqgeqs}) also looks similar to DGLAP expressions:

\begin{eqnarray}\label{fy}
f_{q}(\omega,y) &=& e^{- \omega y}\left[C_{(+)} e^{\Omega_{(+)} y} + C_{(-)} e^{\Omega_{(-)} y}\right],
\\ \nonumber
f_{g}(\omega, y) &=& e^{- \omega y}\left[ C_{(+)} \frac{h_{gg} - h_{qq} + \sqrt{R}}{2h_{qg}} e^{\Omega_{(+)} y} +
C_{(-)} \frac{h_{gg} - h_{qq} - \sqrt{R}}{2h_{qg}} e^{\Omega_{(-)} y} \right].
\end{eqnarray}

This similarity is especially clear as one notices
that the overall factor $e^{- \omega y} = \left(\mu^2/Q^2 \right)^{\omega}$ in Eq.~(\ref{fy}) converts the factor $\left(w/\mu^2\right)^{\omega}$
of Eq.~(\ref{mellin}) into the standard DGLAP factor $x^{- \omega}$, when Eq.~(\ref{fy}) is combined with (\ref{mellin}).
The factors
 $C_{(\pm)} (\omega)$ in Eq.~(\ref{fy}) are arbitrary whereas $\Omega_{(\pm)}$ are expressed through
$h_{ik}$:

\begin{equation}\label{omegapm}
\Omega_{(\pm)} = \frac{1}{2} \left[ h_{gg} + h_{qq} \pm \sqrt{R}\right],
\end{equation}
with

\begin{equation}\label{r}
R = (h_{gg} + h_{qq})^2 - 4(h_{qq}h_{gg} - h_{qg}h_{gq}) = (h_{gg} - h_{qq})^2  + 4 h_{qg}h_{gq}.
\end{equation}

The next step is to specify coefficient functions $C_{(\pm)} (\omega)$
and we notice that similarity of our approach and DGLAP ends at this point.
Indeed, calculating coefficient functions is beyond the scope of DGLAP whereas we continue to apply the IREE approach.
 Before doing it, let us make use of matching $B_{q,g}$ and amplitudes $\widetilde{B}_{q,g}$
which describe the same process in the kinematics where the external photons are (nearly) on-shell, i.e.
with virtualities $Q^2 \approx \mu^2$. It means that $\widetilde{B}_{q,g}$ do not depend on $y$.
It worth mentioning that our strategy here is to some extent similar to the one of
the BFKL-induced models where the BFKL Pomeron is used as an input.
In the $\omega$-space the matching is

\begin{equation}\label{match}
  f_q (\omega,y)|_{y = 0} = \widetilde{f}_q,~~ f_g (\omega,y)|_{y = 0} = \widetilde{f}_g,
\end{equation}
where $\widetilde{f}_{q,g}$ are related by the Mellin transform (\ref{mellin}) to
amplitudes  $\widetilde{B}_{q,g}$.

Combining Eqs.~(\ref{match}) and (\ref{fy}) lead us to the
algebraic system:

\begin{eqnarray}\label{cpmeq}
\widetilde{f}_ q &=& C_{(+)} + C_{(-)},
\\ \nonumber
\widetilde{f}_g &=& C_{(+)} \frac{h_{gg}- h_{qq} + \sqrt{R}}{2 h_{gq}} + C_{(-)} \frac{h_{gg}- h_{qq} - \sqrt{R}}{2 h_{gq}}{2h_{qg}}  ,
\end{eqnarray}
which makes possible to express $C_{(\pm)}$ through $\widetilde{f}_ {1,2}$:

\begin{eqnarray}\label{cpmfeq}
C_{(+)} &=&  \frac{- \widetilde{f}_q \left(h_{gg}- h_{qq} - \sqrt{R}\right) + \widetilde{f}_g 2 h_{qg}}{2\sqrt{R}},
\\ \nonumber
C_{(-)} &=&  \frac{ \widetilde{f}_q \left(h_{gg}- h_{qq} + \sqrt{R}\right) - \widetilde{f}_g 2 h_{qg}}{2\sqrt{R}}.
\end{eqnarray}

Now we have  to calculate $\widetilde{f}_{q,g}$. We do it with constructing and solving appropriate IREEs.
These IREEs   are

\begin{eqnarray}\label{tildefqgeqs}
\omega \widetilde{f}_q (\omega) &=& g_q + h_{qq}(\omega) \widetilde{f}_q (\omega) + h_{gq}(\omega)) \widetilde{f}_g,
 \\ \nonumber
\omega  \widetilde{f}_g (\omega) &=& g_g + h_{qg}(\omega) \widetilde{f}_q (\omega) + h_{gg}(\omega) \widetilde{f}_g (\omega),
 \end{eqnarray}
where inhomogeneous terms $g_{q,g}$ stand for the inputs. We remind that, by definition, the inputs cannot be obtained with
evolving some simpler objects. We will specify $g_{q,g}$ in the next Sect.
 Solution to Eq.~(\ref{tildefqgeqs}) is

\begin{eqnarray}\label{tildefqg}
	\widetilde{f}_q &=& \frac{- g_q (h_{gg}- \omega) + g_g h_{qg}}{\Delta},
	\\ \nonumber
	\widetilde{f}_g &=& \frac{ g_q a_{gq} - g_q (h_{qq}-\omega)}{\Delta},
\end{eqnarray}
where

\begin{equation}\label{delta}
\Delta = (\omega - h_{qq})(\omega - h_{qq}) - h_{qg}h_{gq}.
\end{equation}

Substituting (\ref{tildefqg}) in (\ref{cpmfeq}) allows us to represent $C_{(\pm)}$
through $h_{ik}$ and inputs $g_{q,g}$. We write $C_{(\pm)}$
in the following form:

\begin{eqnarray}\label{cpmg}
C_{(+)} = g_q G^{(+)}_q + g_g  G^{(+)}_g,
\\ \nonumber
C_{(-)} = g_q G^{(-)}_q + g_g  G^{(-)}_g,
\end{eqnarray}
where

\begin{eqnarray}\label{gpm}
 G^{(+)}_q &=& \frac{(h_{qg} - \omega)\left(h_{gg} - h_{qq} - \sqrt{R}\right) + 2 h_{qg}h_{gq}}{2 \Delta \sqrt{R}},
\\ \nonumber
G^{(+)}_g &=&  \frac{- h_{qg}\left(h_{gg} - h_{qq} - \sqrt{R}\right) - 2 h_{qg} (h_{qq} - \omega)}{2 \Delta \sqrt{R}},
\\ \nonumber
 G^{(-)}_q &=& \frac{-(h_{qq} - \omega)\left(h_{gg} - h_{qq} + \sqrt{R}\right) - 2 h_{qg}h_{gq}}{2 \Delta \sqrt{R}},
 \\ \nonumber
G^{(-)}_g &=&  \frac{h_{qg}\left(h_{gg} - h_{qq} + \sqrt{R}\right) + 2 h_{qg} (h_{qq} - \omega)}{2 \Delta \sqrt{R}}.
\end{eqnarray}

Combining Eqs.~(\ref{gpm}), (\ref{cpmg}) and (\ref{fy})  leads to expressions for  $f_{q,g}$ in terms of $h_{ik}$ and $g_{q,g}$.
We remind that explicit expressions for $h_{ik}$ can be found in Appendix B. They are known in DLA for both spin-dependent DIS structure function $g_1$
(see Ref.~\cite{egtg1sum}) and for $F_1$ as well (see Ref.~\cite{etf1}).
Let us compare Eq.~(\ref{tildefqgeqs}) for $\widetilde{f}_{q,g}(\omega)$ and
 Eq.~(\ref{fqgeqs}) for $f_{q,g}(\omega, y)$. The first difference between them is that Eq.~(\ref{tildefqgeqs})
 does not contain the derivative $\partial/\partial y$ because $\widetilde{f}_{q,g}$ do not depend on $y$.
The second difference
is the presence of inhomogeneous terms $g_q$ and $g_g$
 in (\ref{tildefqgeqs}).
 These terms stand for the inputs, i.e. for the starting point of the evolution. Specifying them is necessary for obtaining
 explicit expressions for $f_{q,g}$.
 Below we  consider this issue in detail.

 \section{Specifying inputs $g_q$ and $g_g$ for amplitudes  $B_{q,g}$}

Specifying inputs $g_q$ and $g_g$ is the key point of our paper because it is here that we deviate from the
routine IREE technology. We remind that throughout the history of the
IREE approach the inputs have always been defined as the Born
contributions whereas contributions of  higher loops were obtained with evolving the Born amplitudes.
However, this technology
cannot apply to calculating amplitudes $B_{q,g}$. Indeed,  the Born values for both $B_q$ and $B_g$ are zeros, so
substituting  them in Eq.~(\ref{tildefqgeqs}) would lead to the system of algebraic  homogeneous
equations without an unambiguous  solution. The next option is to choose the first-loop amplitudes as the inputs.
Technically it is possible: they are non-zero (see Eq.~(\ref{fl1})) and evolving them one can obtain  $B_{q,g}$ in DLA.
However, in this case the important second-loop contributions $B^{(2)}_q$ and $B^{(2)}_g$, each $\sim 1/x$
(see Eqs.~(\ref{b2qlead},\ref{b2glead})),
would be left unaccounted because the IR-evolution controls logarithms and cannot generate the factors $1/x$.
In Sect.~2C we presented the scenario
where $B^{(2)}_q$ and $B^{(2)}_g$ were chosen as the inputs and demonstrated that higher loops
cannot change this factor. Instead, they can generate DL contributions which are the most important at small $x$.
Now we implement this scenario in IREEs and choose $B^{(2)}_{q,g}$ as the inputs. To this end, we should
express $B^{(2)}_{q,g}$ in the $\omega$-space.
In the first place we represent $B^{(2)}_{q,g}$ in the following form:

\begin{eqnarray}\label{bqgx}
B^{(2)}_q = \rho \widetilde{B}^{(2)}_q
\\ \nonumber
B^{(2)}_g = \rho \widetilde{B}^{(2)}_g,
\end{eqnarray}
with $\rho = \ln (w/\mu^2)$ (see Eq.~(\ref{lambda})).  Notice that $\rho$ corresponds to $1/\omega^2$ in the $\omega$
space (see Eq.~(\ref{mellin})). Then,
remembering that the Mellin transform does not affect
$1/x$, we write $B^{(2)}_{q,g}$ in the $\omega$-space and obtain
the Mellin amplitudes $\varphi_{q,g}$ conjugated to $B^{(2)}_{q,g}$:

\begin{eqnarray}\label{phiqg}
\varphi_q =  \frac{\widetilde{B}^{(2)}_q}{\omega^2} = \left(\frac{\gamma^{(2)} C^{(2)}_q}{x}\right) \frac{1}{\omega^2}
\equiv \frac{ \gamma^{(2)} b^{(2)}_q (\omega)}{x},
\\ \nonumber
\varphi_g = \frac{\widetilde{B}^{(2)}_g}{\omega^2} = \left(\frac{\gamma^{(2)} C^{(2)}_g}{x}\right) \frac{1}{\omega^2}
\equiv \frac{ \gamma^{(2)} b^{(2)}_g (\omega)}{x} .
\end{eqnarray}

Thus we can fix the inputs $g_q$ and $g_g$ in Eq.~(\ref{cpmg}):

\begin{eqnarray}\label{gbqg}
g_q =  \varphi_q = \gamma^{(2)} b^{(2)}_q/x,
\\ \nonumber
g_g = \varphi_g =  \gamma^{(2)} b^{(2)}_g/x.
\end{eqnarray}

We remind that choosing these inputs  takes us out of the standard form of DLA, where Born amplitudes
were considered as the starting point of evolution.

\section{ Explicit expressions for $F_L$ in DLA}

Substituting $g_{q,g}$ of Eq.~(\ref{gbqg}) in (\ref{tildefqg}) and combining the result with
Eqs.~(\ref{cpmfeq},\ref{fy},\ref{mellin}), we obtain explicit expressions for $B_{q,g}$.
Then, using Eq.~(\ref{flqgdef}) drives us to expressions for  $F_L^{(q,g)}$.
As the obtained expressions are linear in $g_{q,g}$, we can
factorize from them the overall factor $\gamma^{(2)}/x$.
To this end  we introduce $C^{\prime}_{{\pm}}$:

\begin{equation}\label{cpmprime}
C_{{\pm}} = \gamma^{(2)} x^{-1}C^{\prime}_{{\pm}}.
\end{equation}

Using Eq.~(\ref{cpmprime}) allows us to represent expressions for $F_L^{(q,g)}$ as follows:

\begin{eqnarray}\label{flqg}
F_L^{(q)} &=&  4 x~ \gamma^{(2)} \int_{- \imath \infty}^{\imath \infty} \frac{d \omega}{2 \pi \imath} x^{- \omega}
\left[C^{\prime}_{(+)} e^{\Omega_{(+)} y} + C^{\prime}_{(-)} e^{\Omega_{(-)} y}\right],
\\ \nonumber
F_L^{(g)}&=&  4 x~ \gamma^{(2)} \int_{- \imath \infty}^{\imath \infty} \frac{d \omega}{2 \pi \imath} x^{- \omega}
\left[ C^{\prime}_{(+)} \frac{h_{gg} - h_{qq} + \sqrt{R}}{2h_{qg}} e^{\Omega_{(+)} y} +
C^{\prime}_{(-)} \frac{h_{gg} - h_{qq} - \sqrt{R}}{2h_{qg}} e^{\Omega_{(-)} y} \right].
\end{eqnarray}

The overall factor $4x$ at Eq.~(\ref{flqg}) is the product of the
factor $4x^2$ of Eq.~(\ref{flqgdef}) and the factor $1/x$ from the inputs $g_{q,g}$.
 Eq.~(\ref{flqg}) includes the contributions to $F_L^{(q,g)}$
most essential at small $x$.
It does not include the first-loop contribution (\ref{fl1}) and other contributions decreasing at small $x$.
On the contrary, both $F_L^{(q)}$ and $F_L^{(g)}$ of Eq.~(\ref{flqg}) rise
when $x$ is decreasing, albeit this does not look obvious. In order to make it seen clearly we consider below
the small-$x$ asymptotics of $F_L^{(q,g)}$, which look much simpler than the parent
expressions in Eq.~(\ref{flqg}).


\subsection{Small-$x$ asymptotics of $F_L$}

At $x \to 0$, $F_L^{(q,g)}$ can be approximated by their small-$x$
asymptotics which we denote $\left(F_L^{(q,g)}\right)_{AS}$. Technology of calculating the asymptotics is based on the saddle-point
method and the whole procedure is identical to the one for $F_1$. So, we can use
the appropriate results of Ref.~\cite{etf1}. After the asymptotics of $F_L^{(q,g)}$ have been
calculated and convoluted with the parton distributions $\Phi_{q,g}$ (see Eq.~(\ref{fact})),
the small-$x$ asymptotics of $F_L$ is obtained:

\begin{eqnarray}\label{as}
\left(F_L\right)_{AS} \sim \frac{\Pi}{\ln^{1/2}(1/x)} x^{1 -\omega_0} \left(Q^2/\mu^2\right)^{\omega_0/2},
\end{eqnarray}
where the factor $\Pi$ includes both numerical factors of perturbative origin and values of the quark and gluon distributions in the
$\omega$-space at $\omega = \omega_0$. In any form of QCD factorization $\Pi$ does not contain any dependency on $Q^2$ or $x$
(see \cite{etf1} for detail).
Then,  $\omega_0$ is  the Pomeron intercept calculated with DL accuracy. This intercept was
first calculated in Ref.~\cite{etf1}.
We remind that it has nothing in common with the BFKL intercept. It is convenient to represent $\omega_0$ as follows:

\begin{equation}\label{omegad}
\omega_0 = 1 + \Delta^{(DL)}.
\end{equation}

 Numerical estimates for $\Delta^{(DL)}$
depend on accuracy of calculations. When $\alpha_s$ is assumed to be fixed\footnote{we use here the value $\alpha_s = 0.24$ according to prescription of Ref.~\cite{egtalpha}},

\begin{equation}\label{deltafix}
\Delta^{(DL)}_{fix}  = 0.29
\end{equation}
  and

\begin{equation}\label{deltadl}
\Delta^{(DL)} = 0.07,
\end{equation}
when  the $\alpha_s$ running effects are accounted for. Substituting either (\ref{deltafix}) or (\ref{deltadl})
in Eq.~(\ref{as}), one easily finds that $F_L \sim x^{-\Delta^{(DL)}}$ at $x \to 0$.
The asymptotics of $F_1$ was calculated in Ref.~\cite{etf1} showed that asymptotically $F_1 \sim x^{- \omega_0}$
and therefore $F_L \sim 2x F_1$.

The growth of $F_L$ and $x F_1$ at small $x$ is caused by the Pomeron
behaviour of the parton-parton amplitudes $f_{ik} = 8 \pi^2 h_{ik} \sim x^{- \omega_0}$. Amplitudes $f_{gg}$ and $f_{gq}$, being
convoluted with $\Phi_g$ and  $\Phi_q$,  form the gluon distribution in the initial
hadron,  which we denote $G_h$:

\begin{equation}\label{gh}
G_h = h_{gg}\otimes \Phi_g + h_{gq}\otimes \Phi_q.
\end{equation}

So, at small $x$

\begin{equation}\label{flgas}
F_L \sim x G_h.
\end{equation}

 Another interesting observation
following from Eq.~(\ref{as}) is that

\begin{equation}\label{logfl}
2 \frac{\partial \ln F_L}{\partial \ln Q^2} +  \frac{\partial \ln F_L}{\partial \ln x} \to 1
\end{equation}
at $x \to 0$.
We think that it would be interesting to check this relation with analysis of available experimental data. To
conclude discussion of the asymptotics, we notice that the asymptotics (\ref{as}) should be used within its applicability region, otherwise one
should use the expressions of Eq.~(\ref{flqg}). The estimate obtained in Ref.~\cite{etf1}
states that (\ref{as}) can be used at $x \leq 10^{-6}$.

\subsection{Comparison with approaches involving BFKL}

Let us start this comparison with considering the second-order graphs (b) and (c) in Fig.~2,
each with a pair of virtual gluons propagating in the $t$-channel.
In Sect.~III we used the DL
configuration, where one of the gluons is longitudinally polarized while polarization of the other
gluon is transverse
In contrast, contributions to BFKL coming from these graphs involve the kinematics where the both ladder gluons bear
 longitudinal polarizations. Accounting for these polarizations immediately leads to the following behavior of contributions
 $B^{(2b)}_{LL}$  and  $B^{(2c)}_{LL}$  (the subscripts $LL$ refer to the longitudinal polarizations):

\begin{equation}\label{bcll}
B^{(2b)}_{LL} \sim B^{(2c)}_{LL} \sim \frac{1}{x \lambda},
\end{equation}
with $\lambda = \mu^2/w$.
Therefore,  $B^{(2b,2c)}_{LL}$
are greater than the considered in Sec.~III contributions $B^{(2)}_{q,g}$
(we remind that $B^{(2)}_{q,g} \sim  1/x$). Convoluting any of graphs (b,c) in Fig.~2
with a hadron and using appropriate hadron impact factors turns the factor $1/\lambda$ into $1/x$, so the singular
factor in Eq.~(\ref{bcll}) is now $1/x^2$. This factor cancels the factor $x^2$ relating $B^{(2b,2c)}_{LL}$
to $F_L$ (see Eq.~(\ref{flqgdef})).

 Then,
accounting for the impact of higher loops brings the Regge factor $x^{- \Delta_{BFKL} }$,
with $\Delta_{BFKL}$ being the intercept of the BFKL Pomeron. Thus we obtain that the
BFKL contribution to $F_L$ is

\begin{equation}\label{asbfkl}
F^{BFKL}_L \sim x^{- \Delta_{BFKL}},
\end{equation}
where $\Delta_{BFKL}$ is used in either LO or NLO. where $\Delta_{BFKL}$ is used in either LO or NLO.
In contrast to Eq.~(\ref{asbfkl}), the
contribution (\ref{as})  has the extra factor $x$ and because of it (\ref{asbfkl})
may look more important (\ref{as}).
However, the leading singularity $\omega_0$ in
Eq.~(\ref{as}) is large, $\omega_0 > 1$ (see Eq.~(\ref{omegad})), so it cancels the factor $x$ and after that
$F_L \sim x^{- \Delta_{DL}}$. Thus the small-$x$ behaviour of $F_L$ predicted by Eq.~(\ref{as}), and the one predicted by
Eq.~(\ref{asbfkl}) become very much alike.
Indeed, the intercepts of Pomerons in the both approaches are pretty close to each other:
$\Delta^{(DL)}_{fix}$ of Eq.~(\ref{deltafix})
is close to the intercept of the LO BFKL Pomeron and
$\Delta^{(DL)}$ of Eq.~(\ref{deltadl}) practically coincides with the NLO BFKL Pomeron intercept.

On the contrary,
the $Q^2$-dependence predicted by  Eq.~(\ref{as}) differs from predictions given by all other approaches:
they do not satisfy Eq.~(\ref{logfl}).
It means that studying the $x$-dependence of experimental data for $F_L$ with using Regge fits
cannot unambiguously deduce which of the two Pomerons is involved. In order to clear this issue, one should
investigate the $Q^2$-dependence of the data.
To conclude this Section, we once more stress that our approach and the ones involving
BFKL deal with different logarithmic
contributions and cannot be related to each other.

\subsection{Remark on $F_L$ at arbitrary $Q^2$}

The expressions in Eq.~(\ref{flqg}) are valid in the kinematic region (\ref{kin}) where $Q^2$ is large. However, it is easy to generalize Eq.~(\ref{flqg})
to small $Q^2$. It was proved in Refs.~\cite{etf1, egtg1sum} and used for the structute function $F_1$
in Ref.~\cite{etcomb} that such a generalization is achieved with replacement of $Q^2$
by $Q^2 + \mu^2$. When this shift has been done, $F_L^{(q)}$ and  $F_L^{(q)}$ of Eq.~(\ref{flqg})
depend on new variables $\bar{x}, \bar{Q}^2$:

\begin{equation}\label{qbar}
\bar{Q}^2 = Q^2 + \mu^2,~~~\bar{x} = \bar{Q}^2 /w.
\end{equation}

Thus, one can universally use the expressions for $F_L^{(q,g)}$ in Eq.~(\ref{flqg})
at arbitrary $Q^2$ providing the arguments of $F_L^{(q,g)}$  are $\bar{x}$ and $\bar{Q}^2$.

\section{Conclusions}

Our results predict that $F_L$ grows at small $x$ despite the very small factor $x^2$ at $B$ in Eq.~(\ref{fldef}).
First,
we re-calculated with logarithmic accuracy the available in the literature
second-loop contributions  $B^{(2)}_q$ and $B^{(2)}_g$, each contains the large
power factor $1/x$ in contrast to the Born and first-loop contributions. This calculation
allowed us to conclude
that $1/x$  will be present in higher-loop expressions and
cannot disappear or be
replaced by another power factor. We demonstrated
that most important contributions coming from higher orders are
double logarithms. Accounting for DL contributions to all orders in $\alpha_s$,
we calculated
the $x$ and $Q^2$ -evolution of $B^{(2)}_{q,g}$ in DLA. This
evolution proved to be similar to the evolution of the structure function $F_1$. Eventually we obtained
Eq.~(\ref{flqg}) for the partonic components $F_L^{(q)}$ and  $F_L^{(q)}$
of $F_L$. The both these components rise at small $x$ though complexity of
expressions in Eq.~(\ref{flqg}) prevents to see the rise. To make the rise be clearly seen,
we calculated the small-$x$ asymptotics of $F_L$, which proved to be of the Regge type. The asymptotics make obvious that
the synergic effect of the factor $1/x$ and the total resummation of double
logarithms overcomes smallness of the factor $x^2$ at $B$ in Eq.~(\ref{fldef}) and ensures the rise of $F_L$ at small $x$,
see Eq.~(\ref{as}). Then in Eq.~(\ref{flgas}) we noticed that the rise of $F_L$ and the gluon distributions in the hadrons at small $x$ are
identical. We also suggested in Eq.~(\ref{logfl}) the simple relation between
derivatives of logarithm of $F_L$. This relation could be checked with analysis of experimental data, so such check could test correctness of our reasoning. Comparing our results on the asymptotics  of $F_L$ and the ones based on BFKL,
we demonstrated that they predict the similar small-$x$ behavior and widely different $Q^2$-dependence.
The explicit expressions for $F_L$ obtained in Sect.~V  are valid at $Q^2 \geq \mu^2$.  In Sect.~VI we obtained the extension of  those expressions to the region $Q^2 < \mu^2$.
Confronting our results on the asymptotics  of $F_L$ with the ones based on BFKL Pomeron,
we demonstrated that they predicted the similar small-$x$ behavior of $F_L$  and widely different $Q^2$-dependence.

\section{Acknowledgement}

We are grateful to V.~Bertone, N.Ya.~Ivanov and Yuri V.~Kovchegov  for useful communications.

\section{Appendix}

\subsection{Integration in Eq.~(\ref{b2qsud}) }

We write Eq.~(\ref{b2qsud}) in the following form:

\begin{equation} \label{4aidef}
B^{{2a}} \approx  4 C^2_F \chi_2  \left[I^{{2a}}_1 + I^{{2a}}_2\right],
\end{equation}
with $I^{{2a}}_{1,2}$ defined as integrals over the transverse momenta $z_1$:

\begin{eqnarray}\label{4aij}
I^{{2a}}_1 &=& \int^1_{\lambda} \frac{dz_1}{z_1}  J^{{2a}}_1,
\\ \nonumber
I^{{2a}}_2 &=& \int^1_{\lambda} \frac{dz_1}{z_1}   J^{{2a}}_2,
\end{eqnarray}
where $J^{{2a}}_{1,2}$ involve integration over $z_2$:

\begin{eqnarray}\label{4ajjtilde}
	J^{{2a}}_1 &=& \int^1_{\lambda} dz_2 \frac{z^3}{z^2_2}  ~\widetilde{J}^{{2a}}_1,
	\\ \nonumber
	I^{{2a}}_2 &=& \int^1_{\lambda} dz_2 \frac{z^2}{z^2_2} \widetilde{J}^{{2a}}_2.
\end{eqnarray}

Integrals $\widetilde{J}^{{2a}}_{1,2}$ deal with integration over the longitudinal variable $l$:

\begin{eqnarray}\label{4ajl}
\widetilde{J}^{{2a}}_1 &=& -\int_z^1 \frac{d l}{ l^2 (l + \eta)^2 },
\\ \nonumber
\widetilde{J}^{{2a}}_2 &=& \int_z^1 \frac{d l}{ l (l + \eta)^2 },
\end{eqnarray}
with $\eta$ defined in Eq.~(\ref{eta}).
Integration over $l$ in (\ref{4ajl}) yields

\begin{eqnarray}\label{4ajtilde}
\widetilde{J}^{{2a}}_1 &=&
\frac{1}{\eta^2} \left(1 - \frac{1}{z}\right) - \frac{2}{\eta^3} \ln \left( \frac{1 + \eta}{z + \eta}\right)
+ \frac{1}{\eta^3} \left[ \frac{1}{1 + \eta} - \frac{1}{z + \eta}\right],
\\ \nonumber
\widetilde{J}^{{2a}}_2 &=&
\frac{1}{\eta^2} \left[- \ln (1 + \eta) - \ln \left((z + \eta )/z\right) + \frac{\eta}{1 + \eta} - \frac{\eta}{z + \eta} \right]
\end{eqnarray}

and therefore

\begin{eqnarray}\label{4ajint}
J^{{2a}}_1 &=& \int_{\lambda}^1 dz_2
\left[\frac{z - 1}{(z_2 + x)^2} - \frac{z_2}{(z_2 + x)^3}\ln U(z,z_2)
\right.  \\ \nonumber  &~& \left.
+ \frac{z_2}{(z_2 + x)^3}\ln (2 z_2 + x)
+ \frac{z_2}{(z_2 + x)^3U(z,z_2)} - \frac{z_2}{(z_2 + x)^3 (2 z_2 + x)}
\right]
\\ \nonumber
J^{{2a}}_2 &=& \int_{\lambda}^1 dz_2
\frac{1}{(z_2 + x)^2}
\left[ \ln (2 z_2 + x) - \ln U(z,z_2) - \frac{z_2}{U(z,z_2)} + \frac{z_2}{2 z_2 + x}\right],
\end{eqnarray}
where

\begin{equation}\label{u}
U(z,z_2) = z_2 + z (z_2 + x).
\end{equation}

It is convenient to perform integration in (\ref{4ajint}), using the variable $y = 1/(z_2 + x)$ instead of $z_2$.
The most essential
contributions in (\ref{4ajint}) at small $x$ are the ones $\sim 1/x$. Accounting for them only, we obtain

\begin{eqnarray}\label{4aj}
J^{{2a}}_1 &=& x^{-1} [\ln 2 - 1/2],
\\ \nonumber
 J^{{2a}}_2 &=& x^{-1} (1/2) .
\end{eqnarray}

Substituting this result in (\ref{4aij}), we obtain

\begin{equation}\label{4ai}
I^{{2a}}_1 + I^{{2a}}_2  = (\rho \ln2) ~x^{-1}.
\end{equation}

\subsection{Expressions for $h_{ik}$}

\begin{eqnarray}\label{h}
&& h_{qq} = \frac{1}{2} \Big[ \omega - Z - \frac{b_{gg} -
b_{qq}}{Z}\Big],\qquad h_{qg} = \frac{b_{qg}}{Z}~, \\ \nonumber &&
h_{gg} = \frac{1}{2} \Big[ \omega - Z + \frac{b_{gg} -
b_{qq}}{Z}\Big],\qquad h_{gq} =\frac{b_{gq}}{Z}~,
\end{eqnarray}
where
\begin{equation}
\label{z}
 Z = \frac{1}{\sqrt{2}}\sqrt{ Y + W
}~,
\end{equation}
with
\begin{equation}\label{y}
Y = \omega^2 - 2(b_{qq} + b_{gg})
\end{equation}
and
\begin{equation}\label{w}
  W = \sqrt{(\omega^2 - 2(b_{qq} + b_{gg}))^2 - 4 (b_{qq} - b_{gg})^2 -
16b_{gq} b_{qg} },
\end{equation}

where the terms $b_{rr'}$ include the Born factors $a_{rr'}$ and contributions of non-ladder graphs $V_{rr'}$:
\begin{equation}\label{bik}
b_{rr'} = a_{rr'} + V_{rr'}.
\end{equation}

The Born factors are (see Ref.~\cite{egtg1sum} for detail):

\begin{equation}\label{app}
a_{qq} = \frac{A(\omega)C_F}{2\pi},~a_{qg} = \frac{A'(\omega)C_F}{\pi},~a_{gq} = -\frac{A'(\omega)n_f}{2 \pi}.
~a_{gg} = \frac{2N A(\omega)}{\pi},
\end{equation}
where $A$ and $A'$ stand for the running QCD couplings as shown in Ref.~\cite{egtalpha}:

\begin{eqnarray}\label{a}
A = \frac{1}{b} \left[\frac{\eta}{\eta^2 + \pi^2} - \int_0^{\infty} \frac{d z e^{- \omega z}}{(z + \eta)^2 + \pi^2}\right],
A' = \frac{1}{b} \left[\frac{1}{\eta} - \int_0^{\infty} \frac{d z e^{- \omega z}}{(z + \eta)^2}\right],
\end{eqnarray}
with $\eta = \ln \left(\mu^2/\Lambda^2_{QCD}\right)$ and $b$ being the first coefficient of the Gell-Mann- Low function. When the running effects for the QCD coupling
are neglected,
$A(\omega)$ and $A'(\omega)$ are replaced by $\alpha_s$.
The terms $V_{rr'}$ approximately represent the impact of non-ladder graphs on $h_{rr'}$ (see Ref.~\cite{egtg1sum} for detail):

\begin{equation}
\label{vik} V_{rr'} = \frac{m_{rr'}}{\pi^2} D(\omega)~,
\end{equation}
with
\begin{equation}
\label{mik} m_{qq} = \frac{C_F}{2 N}~,\quad m_{gg} = - 2N^2~,\quad
m_{gq} = n_f \frac{N}{2}~,\quad m_{qg} = - N C_F~,
\end{equation}
and
\begin{equation}
\label{d} D(\omega) = \frac{1}{2 b^2} \int_{0}^{\infty} d z
e^{- \omega z} \ln \big( (z + \eta)/\eta \big) \Big[
\frac{z + \eta}{(z + \eta)^2 + \pi^2} - \frac{1}{z +
\eta}\Big]~.
\end{equation}

Let us note that $D = 0$ when the running coupling effects are neglected. It corresponds the total compensation of DL
contributions of non-ladder Feynman graphs to scattering amplitudes with the positive signature as was first
noticed in Ref.~\cite{nest}. When $\alpha_s$ is running, such compensation is only partial.

\end{document}